\documentclass[twocolumn,letterpaper,showpacs,preprintnumbers,amsmath,prl,floatfix]{revtex4}

\usepackage{graphicx}
\usepackage{bm}

\begin{document}

\title{Generation of anomalously energetic suprathermal electrons by an
electron beam interacting with a nonuniform plasma}

\author{D. Sydorenko$^{a}$}
\author{I. D. Kaganovich$^{b}$}
\author{L. Chen$^{c}$}
\author{P. L. G. Ventzek$^{c}$}

\affiliation{
    $^{a}$University of Alberta, Edmonton, Alberta T6G 2E1, Canada \\
    $^{b}$Princeton Plasma Physics Laboratory, Princeton University,
          Princeton, New Jersey 08543, USA \\
    $^{c}$Tokyo Electron America, Austin, Texas 78741, USA}

\begin{abstract}

Generation of anomalously energetic suprathermal electrons was observed in
simulation of a high-voltage dc discharge with electron emission from the
cathode. An electron beam produced by the emission interacts with the
nonuniform plasma in the discharge via a two-stream instability. Efficient
energy transfer from the beam to the plasma electrons is ensured by the plasma
nonuniformity. The electron beam excites plasma waves whose wavelength and
phase speed gradually decrease towards anode. The short waves near the anode
accelerate plasma bulk electrons to suprathermal energies. The sheath near the
anode reflects some of the accelerated electrons back into the plasma. These
electrons travel through the plasma, reflect near the cathode, and enter the
accelerating area again but with a higher energy than before. Such particles
are accelerated to energies much higher than after the first acceleration. This
mechanism plays a role in explaining earlier experimental observations of
energetic suprathermal electrons in similar discharges.

\end{abstract}




\maketitle

The collisionless relaxation of high-current electron beams and plasma jets is
important for many applications: solar disruptions \cite{MelroseBook1986},
collective stopping of intense electron beams for inertial fusion applications
\cite{MalkinPRL2002,KempPRL2006}, collisionless shocks and collision of flowing
plasmas in astrophysics
\cite{KeloggPSS2003,TreumannAAR2009,BaloghISSISRS2011,MoritaJPCS2010,RossPP2012},
generation of suprathermal electrons \cite{YoonPRL2005}. Such beams are common
in laboratory plasmas. Electrons emitted by electrodes surrounding or immersed
in the plasma are accelerated by the sheath electric field and become the
electron beams penetrating the plasma. In plasma applications where controlling
the electron velocity distribution function (EVDF) is crucial
\cite{KaganovichPPCF2009}, these beams are an important factor capable of
modifying the EVDF and affecting the discharge properties.

Recently, Xu \textit{et al}.~\cite{XuAPL2008} and Chen and Funk
\cite{ChenICPR2010} reported an EVDF measured in a dc-rf discharge with 800 V
dc voltage which has not only a peak at 800 eV corresponding to the electrons
emitted from the dc-biased electrode, but also a significant fraction of
accelerated electrons with energy up to several hundreds eV. In
Refs.~\cite{XuAPL2008,ChenICPR2010}, the acceleration is explained invoking
nonlinear wave-particle interactions
\cite{TsytovichBook1970,ShapiroBPPbook1984,PorkolabRMP1978,RobinsonRMP1997,
CairnsPRL1999,GolovanovSJPP1977,ArzhannikovFT1999}. It is suggested that the
800 eV beam excites fast long plasma waves which decay parametrically into ion
acoustic waves and short plasma waves with much lower phase velocity, then the
short waves accelerate slow electrons of the low-temperature plasma bulk. A
similar mechanism may explain enhanced energetic tails in the energy spectrum
of electrons in aurora \cite{PapadopoulosJGR1974,RowlandJGR1988,YoonPRL2005}.
Other mechanisms invoked to explain the EVDF structure include phase-bunching,
kinematic effect of electrons being trapped between rf and dc potentials and
released towards the rf electrode during short period  of time when rf voltage
is small compared to the dc voltage \cite{KhrabrovPSST2015sub}.

In order to further understand the mechanism of acceleration in the experiment
of Refs.~\cite{XuAPL2008,ChenICPR2010}, a beam-plasma system was simulated
using a 1D3V particle-in-cell code EDIPIC \cite{SydorenkoPhDTh2006}. The EDIPIC
was extensively benchmarked against available analytical predictions for two
stream instability in linear and nonlinear regimes. 
Simulation results discussed below show that the acceleration may be caused not
only by the nonlinear effects, but by the effects related to the plasma
nonuniformity as well. Note that profiling the background plasma density is a
method to control the beam energy deposition in plasma
\cite{AstrelinJETP1998,MalkinPRL2002}. In particular, one can excite the
instability and achieve beam energy deposition in desirable regions while
suppressing the instability in undesirable regions by employing convective
stabilization effects \cite{RyutovJETP1970,GolovanovSJPP1977}. In
Refs.~\cite{MalkinPRL2002,GunellPRL1996,McFarlandPRL1998,WendtPS2001} an
experimental study and particle-in-cell simulations of interaction of a strong,
warm electron beam with an inhomogeneous, bounded plasma showed that the
trapped Langmuir waves can determine the frequencies and areas of localization
of the wave field.
In those studies the beam energy was relatively low (40 eV) and the acceleration of plasma electrons was not observed.

The present paper discusses the mechanism of generation of suprathermal electrons found in the simulation. In this
process, the beam excites long plasma waves in the plasma body, the long plasma waves are converted into short plasma
waves at the plasma periphery where the plasma density decreases, and the short waves accelerate bulk electrons. Some
of these electrons return into the acceleration area again due to reflections from the sheath regions near the plasma
boundaries. The second interaction with the short plasma waves yields electrons with anomalously high energies.
Repeating the acceleration for the same particle is more efficient than a one-time interaction between particles and
plasma waves in density gradient areas considered \textit{e.g.} in
\cite{MelroseBook1986,MalkinPRL2002,GolovanovSJPP1977,WendtPS2001, GunellPRL1996,McFarlandPRL1998}.

\begin{figure}[tbp]
\includegraphics {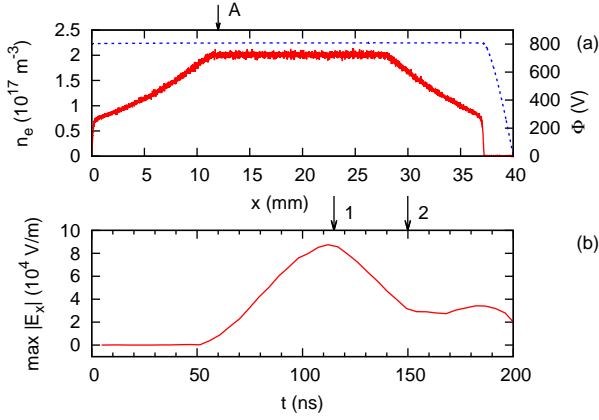}
\caption{\label{fig:01} %
(a) Profiles of electron density (red) and electrostatic potential (blue) immediately before the electron emission at
the cathode begins, $t=49\text{~ns}$. (b) Amplitude of electric field oscillations vs time at point $x=12\text{~mm}$
(marked by arrow A in (a)). Arrows 1 and 2 in (b) mark times when snapshots shown in
Figures~\ref{Evsx_eppmap_evxdf_400Am2_before} and \ref{Evsx_eppmap_evxdf_400Am2_after}, respectively, are obtained. }
\end{figure}

The simulated collisionless argon plasma is bounded between anode (x=0) and cathode (x=40\text{~mm}). The cathode
potential is zero, the anode potential is $800\text{~V}$. The cathode can emit electrons with a constant flux and a
temperature of $1\text{~eV}$. The initial state of the system shown in Fig.~\ref{fig:01} is the result of a preliminary
simulation which starts with uniform plasma density $n_0=2\times 10^{17}\text{~m}^3$, the emission from the cathode
turned off, the electron temperature $T_{e,0}=2\text{~eV}$, the ion temperature $T_{i,0}=0.03\text{~eV}$. The
preliminary simulation lasts for $4000\text{~ns}$. During this time, wide areas with plasma density gradients form as
shown in Fig.~\ref{fig:01}a. The time counter in the simulations with the beam starts at $t=0$ and the electron
emission starts at $t=50\text{~ns}$. The emission current is $20.13\text{~mA/cm}^2$, the relative beam density
$\alpha=n_b/n_0$ in the plasma density plateau area ($11\text{~mm}<x<27\text{~mm}$) is $\alpha=3.77\times 10^{-4}$.
%

\begin{figure}
\includegraphics {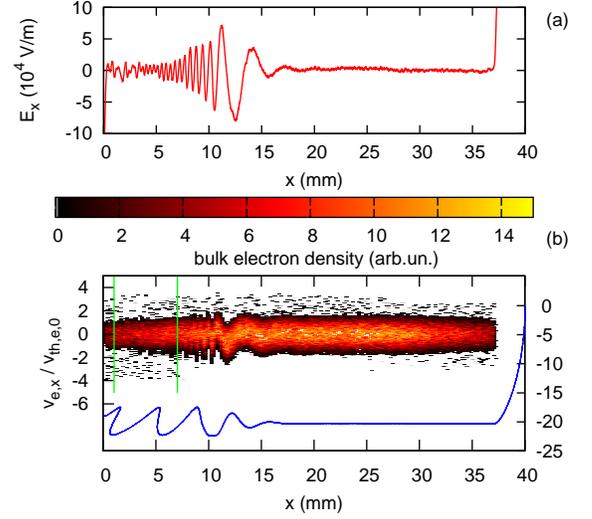}
\caption{\label{Evsx_eppmap_evxdf_400Am2_before} %
(a) Profile of the electric field. (b) Phase plane ``coordinate-velocity'' of bulk electrons (colormap, left vertical
coordinate axis) and the electron beam (blue markers, right vertical coordinate axis). The snapshots are obtained at
time $114.97\text{~ns}$ marked by arrow 1 in Figure~\ref{fig:01}b.}
\end{figure}

\begin{figure}
\includegraphics {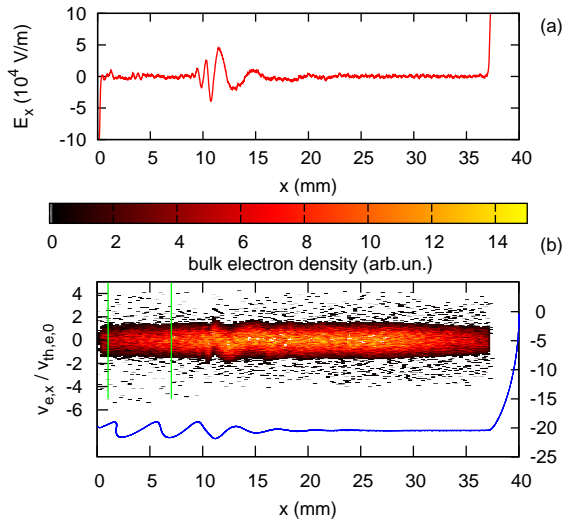}
\caption{\label{Evsx_eppmap_evxdf_400Am2_after} %
Same as in Figure~\ref{Evsx_eppmap_evxdf_400Am2_before}, but the snapshots are obtained at time $149.96\text{~ns}$
marked by arrow 2 in Figure~\ref{fig:01}b.}
\end{figure}

\begin{figure}
\includegraphics {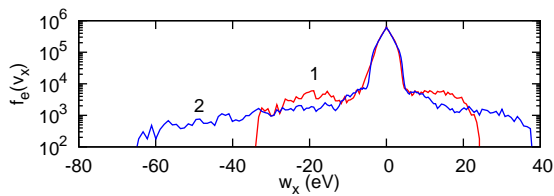}
\caption{\label{evxdf_400Am2_1stage_2stage} %
Bulk electron velocity distribution function obtained using particles with $1\text{~mm}<x<7\text{~mm}$, the horizontal
axis is in energy units, negative values correspond to propagation in the negative $x$-direction. The red and blue
curves are for the system state shown in Figures~\ref{Evsx_eppmap_evxdf_400Am2_before} and
\ref{Evsx_eppmap_evxdf_400Am2_after}, respectively.}
\end{figure}

In the density plateau area, the plasma waves exited by the beam have the wavelength close to $2\pi v_b/\omega_{0,pe}$
where $\omega_{0,pe}=e(n_0/m_e\varepsilon_0)^{1/2}$ and the amplitude growing along the beam propagation towards anode
\cite{KaganovichPRL2015sub}, see Figs.~\ref{Evsx_eppmap_evxdf_400Am2_before}a and
\ref{Evsx_eppmap_evxdf_400Am2_after}a. In the density gradient area, $x<12\text{~mm}$, the profile of the electric
field drastically changes: the amplitude and the wavelength decrease towards anode. Phase plots in
Figs.~\ref{Evsx_eppmap_evxdf_400Am2_before}b and \ref{Evsx_eppmap_evxdf_400Am2_after}b show that bulk plasma and beam
electrons perform strong oscillations in the electric field of the plasma waves. Suprathermal electrons (STEs) are
clearly seen in the EVDF near the anode, see Fig.~\ref{evxdf_400Am2_1stage_2stage}. The electrons are accelerated from
the initial Maxwellian EVDF with the electron temperature $T_{e,0}=2\text{~eV}$ to energies up to 60 eV.
%

\begin{figure}[tbp]
\includegraphics {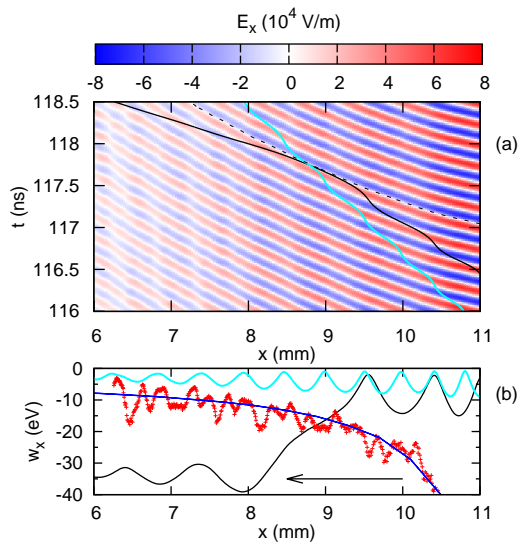}
\caption{\label{fig:tp_Exvst2d_Wevsx_w0_7eV_4eV_reduced} %
Interaction of test particles of different initial energy with the waves. (a) Color map of the electric field as a
function of coordinate and time $E_x(x,t)$. The positive electric field shown with the red color is directed rightward,
towards the cathode. The curves are the test particles trajectories $x(t)$. (b) Energy of test particles vs coordinate.
In (a) and (b), the black solid curves correspond to a particle with initial energy $7~\text{eV}$ at $x=20\text{~mm}$
at $t=110.77\text{~ns}$, the cyan curves correspond to a particle with initial energy $4\text{~eV}$ at $x=20\text{~mm}$
at $t=108.08\text{~ns}$. The solid blue curve in (b) marks the electron energy corresponding to the theoretical local
phase velocity of the wave given by Eq.~\ref{eq:01}. In (b), the negative values of energy correspond to propagation in
the negative $x$ direction (towards the anode), the arrow points in the direction of propagation of the test particles.
}
\end{figure}

\begin{figure}[tbp]
\includegraphics {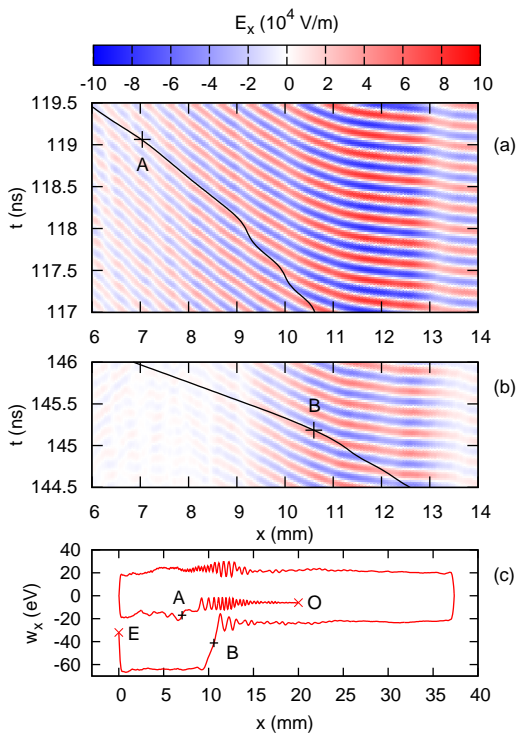}
\caption{\label{fig:tp_45_6eV_recycling} %
Two-stage acceleration of a test particle. The initial energy is $6~\text{eV}$ at $x=20\text{~mm}$ at
$t=110.44\text{~ns}$. (a) The first stage: color map of the electric field as a function of coordinate and time
$E_x(x,t)$. The positive electric field shown with the red color is directed rightward, towards cathode. The curve is
the test particle trajectory $x(t)$. (b) Same as (a) but for the second stage. (c) Energy of the test particle vs
coordinate. In (b), the negative values of energy correspond to propagation in the negative $x$ direction (towards the
anode). Point O marks the starting point, point A -- acceleration in the first stage, point B -- acceleration in the
second stage, and point E -- escape of particle at the anode. }
\end{figure}

The process of acceleration is demonstrated in Figs.~\ref{fig:tp_Exvst2d_Wevsx_w0_7eV_4eV_reduced} and
\ref{fig:tp_45_6eV_recycling} where the color map of the electric field E(t,x) is presented along with the trajectories
of few electrons. The change of the slope of field pattern seen in the color map
(Figs.~\ref{fig:tp_Exvst2d_Wevsx_w0_7eV_4eV_reduced}a and \ref{fig:tp_45_6eV_recycling}a) corresponds to the decrease
of the phase velocity towards the anode due to the plasma density inhomogeneity. The frequency of the wave excited by
the beam is the same everywhere, including the density gradient regions, and is very close to the plasma frequency in
the density plateau $\omega_{0,pe}$. Assuming that the plasma wave satisfies the dispersion relation
$\omega^2=\omega_{pe}^2+3k^2T_e/m_e$ and setting $k=\omega_{0,pe}/v_b$ in the density plateau, the wave frequency is
$\omega_{0,pe}(1+3T_e/m_e v_b^2)^{1/2}$. Substituting this frequency into the dispersion equation and using
$\omega_{pe}^2=\omega_{0,pe}^2n(x)/n_0$, one can obtain the phase velocity of the wave as a function of coordinate
\begin{equation}\label{eq:01}
v_{ph}(x)=\dfrac{\sqrt{3T_e/m_e}} %
                {\sqrt{1-(1-3T_e/m_ev_b^2)n(x)/n_0}}~,
\end{equation}
where $T_e$ and $m_e$ are the electron temperature and mass, $n_e(x)$ is the plasma density which depends on coordinate
$x$, and $v_b$ is the beam velocity. The analytical expression for the phase velocity (\ref{eq:01}) is in good
agreement with the results of the numerical simulation, compare the blue curve with the red markers in
Fig.~\ref{fig:tp_Exvst2d_Wevsx_w0_7eV_4eV_reduced}b.
%
%

After the emission from the cathode started, STEs moving towards the anode appear first downstream of the area with the
strongest oscillations, see Fig.~\ref{Evsx_eppmap_evxdf_400Am2_before}b for $x<8\text{~mm}$. At the peak of the
instability (arrow 1 in Fig.~\ref{fig:01}b), maximal energy of these electrons reaches about $34\text{~eV}$, see curve
1 in Fig.~\ref{evxdf_400Am2_1stage_2stage}. While the accelerated electrons escape to the anode, the plasma potential
relative to the anode increases. Some STEs are reflected by the enhanced potential barrier of the anode sheath and,
after a round trip through the plasma with another reflection near the cathode, they return into the area with intense
plasma oscillations. When it happens, the energies of STEs moving towards the anode downstream of the instability
maximum increase even more, see Fig.~\ref{Evsx_eppmap_evxdf_400Am2_after}b and curve 2 in
Fig.~\ref{evxdf_400Am2_1stage_2stage}. The maximal energy of STEs (registered at $t=139.97\text{~ns}$) is about
$80\text{~eV}$. At the first minimum of the electric field amplitude as a function of time (arrow 2 in
Fig.~\ref{fig:01}b), the maximal energy is lower, about $65\text{~eV}$, but still almost twice the maximal energy of
the one-stage acceleration, compare curves 2 and 1 in Fig.~\ref{evxdf_400Am2_1stage_2stage}.
%
%

Acceleration of bulk electrons by the short waves is studied with test particles which move in the electric field
provided by the simulation. A test particle with low initial energy ($4\text{~eV}$) does not get net acceleration, its
energy only oscillates along the particle trajectory, see cyan curves in
Fig.~\ref{fig:tp_Exvst2d_Wevsx_w0_7eV_4eV_reduced}. A particle with higher initial energy ($7\text{~eV}$), however, is
accelerated by these waves, see the black curves in Fig.~\ref{fig:tp_Exvst2d_Wevsx_w0_7eV_4eV_reduced}. Note that
before and after the acceleration, the energy of the $7\text{~eV}$ test particle in the wave field oscillates, see the
black curve in Fig.~\ref{fig:tp_Exvst2d_Wevsx_w0_7eV_4eV_reduced}b. The acceleration occurs if at the moment when the
energy of the particle is maximal, the velocity of the particle is equal to the local phase speed of the wave, compare
the black energy-vs-coordinate curve of the $7\text{~eV}$ test particle and the blue curve corresponding to the wave
phase velocity in Fig.~\ref{fig:tp_Exvst2d_Wevsx_w0_7eV_4eV_reduced}b.

The average energy of the $7\text{~eV}$ test particle after the acceleration is about $34\text{~eV}$ which is
sufficient to penetrate through the sheath potential barrier (about $23\text{eV}$). Another test particle with initial
energy of $6\text{~eV}$ after the first acceleration (point A in Figs.~\ref{fig:tp_45_6eV_recycling}a and
\ref{fig:tp_45_6eV_recycling}c) has only $20\text{~eV}$. This particle is reflected by the anode sheath, travels
through the whole plasma, reflects near the cathode, and approaches the instability area with the energy of about
$20\text{~eV}$, see Fig.~\ref{fig:tp_45_6eV_recycling}c. The particle is accelerated again at the segment between
$9.5\text{~mm}$ and $11.5\text{~mm}$, see point B in Figs.~\ref{fig:tp_45_6eV_recycling}b and
\ref{fig:tp_45_6eV_recycling}c. The energy of the particle after the acceleration is about $65\text{~eV}$, which is
more than ten times the initial particle energy. The first acceleration occurs similar to the case described in the
previous paragraph except that due to the lower initial particle speed the acceleration occurs closer to the anode
where the wave amplitude is lower, compare Figs.~\ref{fig:tp_45_6eV_recycling}b and
\ref{fig:tp_Exvst2d_Wevsx_w0_7eV_4eV_reduced}a.

The second acceleration occurs in the area immediately adjacent to the density plateau where the wave phase speed and,
in general, amplitude are higher. It is necessary to mention, however, that by this time the short waves in the
majority of the density slope area decayed because they were involved in intense acceleration of plasma particles, see
Figs.~\ref{fig:tp_45_6eV_recycling}b and \ref{Evsx_eppmap_evxdf_400Am2_after}a for $x<9\text{~mm}$, and the instability
amplitude is also noticeably lower than during the first acceleration, compare wave fields in
Figs.~\ref{fig:tp_45_6eV_recycling}a and \ref{fig:tp_45_6eV_recycling}b. The decay of the plasma waves is a significant
limiting factor for the observed two-stage accelerating mechanism.

Summarizing, in a dc plasma-beam system with nonuniform density, suprathermal electrons are generated by short plasma
waves excited at the density gradients. Some of the accelerated electrons may be reflected by the anode sheath and
reintroduced into the two-stream instability area where they will be accelerated one more time. The energy of an
electron after the second acceleration can be an order of magnitude higher than its initial energy and few times more
than its energy after the first acceleration stage.
The simulation above, unlike the experiments of \cite{XuAPL2008,ChenICPR2010}, does not include the rf voltage. This
has both favorable and negative consequences. On one hand, variation of the energy of the beam by the rf voltage may
reduce the efficiency of excitation of plasma waves. On the other hand, in a real dc-rf system, plasma bulk electrons
are trapped (at least part of the rf period) in a potential well much deeper than the one created by thermal electron
motion. This will prevent suprathermal electrons from escaping to the walls and allow them to go through two or more
acceleration cycles as described above.

\section*{ACKNOWLEDGMENTS}

I.~D.~Kaganovich is supported by the U.S. Department of Energy.

%

\end{document}